  \documentclass[conference]{IEEEtran}

   \ifCLASSINFOpdf   
\else
\fi
\usepackage{times,epsfig,latexsym,amssymb,psfrag}
\usepackage{paralist,citesort}
\usepackage{booktabs}
\usepackage{subcaption}
\usepackage{eurosym}
\usepackage{mathrsfs}
\usepackage{enumitem}
\usepackage[ruled,vlined]{algorithm2e}
\usepackage{graphicx}
\usepackage{epstopdf} 
\usepackage{amsmath}
\usepackage{mathtools}

 \newcommand{\qed}{\nobreak \ifvmode \relax \else
 	\ifdim\lastskip<1.5em \hskip-\lastskip
 	\hskip1.5em plus0em minus0.5em \fi \nobreak
 	\vrule height0.75em width0.5em depth0.25em\fi}

\begin{document}
 \title{Self-organized Low-power IoT Networks: A  Distributed Learning Approach}

\author{Amin Azari and Cicek Cavdar\\
KTH Royal Institute of Technology, Email: \{aazari, cavdar\}@kth.se} 
 \maketitle

\begin{abstract}
Enabling large-scale energy-efficient Internet-of-things (IoT) connectivity is an essential step towards realization of networked society. While legacy wide-area wireless systems are highly dependent on network-side coordination, the level of consumed energy in signaling, as well as the expected increase in the number of IoT devices, makes such centralized approaches infeasible in future. Here, we address this problem by   self-coordination for IoT networks through learning from past communications. To this end, we first study low-complexity distributed learning approaches applicable in IoT communications. Then, we present a learning solution to adapt communication parameters of devices to the environment for  maximizing energy efficiency and reliability in data transmissions.  Furthermore, leveraging tools from stochastic geometry, we evaluate the performance of   proposed  distributed learning solution against the centralized coordination. Finally, we analyze the interplay amongst energy efficiency, reliability of communications against noise and interference over data channel, and reliability against adversarial interference over data and feedback channels. The simulation results indicate that compared to the state of the art approaches, both energy efficiency and reliability in IoT communications could be significantly improved using the proposed  learning approach. These promising results, which are achieved using lightweight learning,  make our solution favorable in many low-cost low-power IoT applications.

\end{abstract}
 
\IEEEpeerreviewmaketitle
 
 
 \section{Introduction}
From the first to the fourth generation (4G) of wireless networks, a majority of resources in telecommunications have been dedicated to optimize communication systems with respect to the physical channel constraints, such as noise and  interference \cite{pet}. Thanks to the large communication bandwidth and advanced hardware/software used in both 4G base stations and user devices, 4G networks are able to offer high-speed, seamless, and reliable connectivity to users.  
Compared to the previous generations, the fifth-generation of wireless networks (5G) has an increased focus on providing connectivity for energy/complexity/cost constrained smart devices, i.e. Internet-of-things (IoT) \cite{5g_iot}. The long-term  envision is to provide low-cost, large-scale, and ultra-durable connectivity for everything  which benefits from being connected.   Until now, the design and optimization of  communication networks have been based on statistical models  for arriving traffic as well as physical constraints like noise and  interference. User devices, mostly smart-phones with a daily charging routine,  listen frequently (in the order of sub-seconds) to their serving base stations (BSs), which are responsible for managing the connections, sending connection instructions, and scheduling radio resources.  
As complexity, scale and heterogeneity of wireless networks, especially due to the  IoT traffic, availability of statistical models for arriving traffic in the network-side and ability of energy-limited devices in frequent listening to the access network  become infeasible \cite{datad}. The latter is mainly due to the fact that in IoT networks, the design objectives, quality of service (QoS) requirements, and communications' characteristics are fundamentally different than the ones of legacy communication networks \cite{sysreq}. The state-of-the-art wide area  IoT enabling solutions could be categorized as evolutionary and revolutionary solutions. The former includes solutions implemented by the 3GPP to accommodate IoT traffic in existing cellular infrastructure, e.g. LTE category 1 and M \cite{nok1}. The latter includes solutions which aim at enabling IoT communications in a narrow bandwidth with decreased signaling between devices and the access network \cite{mag_all}. Examples of such solutions are NB-IoT (inside 3GPP), and SigFox and LoRa (outside 3GPP). 

SigFox and LoRa,  the two dominant IoT solutions over the  unlicensed band, benefit from a  simplified connectivity procedure, called grant-free access, which removes need for pairing,  synchronization, and access reservation. Thanks to the reduced signaling in grant-free access, these solutions are able to offer  ultra-long battery lifetimes in IoT communications \cite{lif_com}. Beside solutions over the unlicensed band, the grant-free  access is expected to be also included in future releases of the 3GPP LTE \cite{rsma}.  While energy consumption  in the grant-free radio access mode is extremely low thanks to the removal of signaling procedures, the reliability of communications in this mode is a bottleneck \cite{int2,meysam}.  For example, the ever-increasing coexistence of communications technologies over the ISM band, and lack of dynamic control over operation of IoT devices using these radio resources, result in  no performance guarantee for IoT communications in this band \cite{int2}.    Fig. \ref{suc-opt} represents interference measurements in the European 868 MHz ISM band in Alborg, Denmark \cite{int2}. One sees in this figure that in use-cases like business park, the ISM band suffers from a high level of interference in some sub-bands, which on the other hand,  means  a high probability of collision on them. On the other hand, one sees that in case of smart transmission sub-channel selection, there are sub-bands which suffer from sporadic interference, and hence, probability of success over them is much higher. Similar problem, but in the  code-domain, could be seen in spreading factor distribution of LoRa wide area networks (LoRaWAN), as discussed in \cite{eql}.   
 In recent years, there is an ever increasing interest in leveraging machine  learning tools for characterizing  large-scale networks, where there is no statistical model for describing their behaviors, as well as for operation control of independent nodes which have limited information about their environments and get information only from interactions with their environments  \cite{learn5g}.  In \cite{enb}, network-side reinforcement learning has been proposed for overload control in LTE systems serving massive IoT  traffic. In \cite{adhoc}, self-organized clustering and clustered-access for massive IoT deployments have been investigated.  In \cite{mab7}, use of  multi-arm bandit (MAB) for IoT networks has been proposed, where devices learn how to avoid sub-channels suffering from a high level of static interference. In \cite{seciot}, security concerns coming from connecting IoT devices, i.e. sensors and actuators, to the Internet have been presented, and use of machine learning tools in answering  such concerns has been investigated. 
  To realize self-organized IoT networks able to adapt  themselves to the environment, here we investigate communication in coexistence scenarios, in which the choice of communications parameters, including data rate, sub-channel,  transmit power, and number of repetitions, affects both capacity and battery lifetime of the network. Our aim is to enable low-cost IoT devices to increase reliability of their 
communications, while keeping their energy consumptions as low as possible. In order to investigate application of the derived results in practice, we further present a distributed learning approach for operation control in  LoRa technology, and  compare the results against the analytical results from solving the equivalent centralized optimization problem. The performance evaluation results indicate significant decrease in energy consumption  and increase in probability of success in communications. The main contributions of this work include:
 \begin{itemize}
 \item
 Present a lightweight learning approach designed based on internal and external regret for increasing  energy efficiency and reliability of IoT communications, respectively, with reduced network intervention. 
  \item
 Develop an analytical model for performance evaluation  of the distributed learning solutions by leveraging tools from  stochastic geometry.
 \item
Present distributed learning for operation control of IoT devices utilizing LoRa technology. Evaluate reliability and energy efficiency  of communications utilizing the proposed and benchmark solutions. Highlight tradeoffs between reliability of communications against unintended/adversarial interference and energy efficiency.

 \end{itemize}

The remainder of this paper has been structured as follows. System model is investigated in the next section. The proposed learning approaches are presented in section III. 
In section IV, distributed learning is employed for operation control of LoRa devices, and its performance  is compared against the centralized optimized solution. Simulation results are presented in section V. Concluding remarks are given in section VI.

 \begin{figure}[t!]
        \centering
                \includegraphics[width=3in]{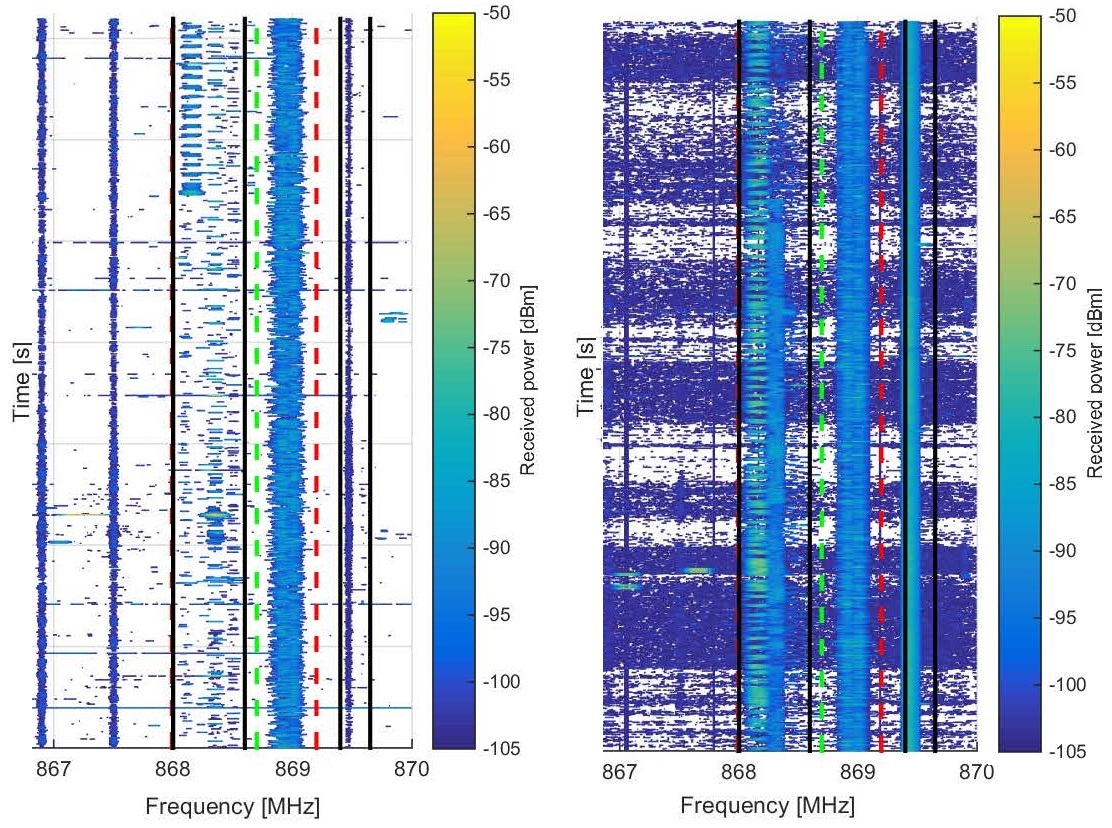}
                 \caption{Interference measurement in the ISM band \cite{int2}. Left: business park, Right: hospital complex. (\copyright 2017 IEEE)}
                \label{int}
\end{figure}

\section{System Model and Problem Formulation}
 A multitude of  IoT devices, denoted by set $\Phi$, have been distributed  in a wide service area. Different IoT devices differ in radio resource usage pattern, i.e. in average reporting period, signal bandwidth, transmit power, and data rate (packet transmission time).  A frequency bandwidth of $\mathcal W$ is shared for  communications, on which the power spectral density of noise is denoted by $\mathcal N$.  We aim at collecting data from a subset of IoT devices\footnote{For example, one may consider coexistence of LoRa and SigFox in an environment, where LoRa receiver treats SigFox signal as interference.}, $\Phi_{\text s} \subset \Phi$, and treat traffic from other devices as interference. 

 The problem to be tackled is  operation control for  devices of interest, i.e. members of $\Phi_{\text s}$, by considering operations of all other devices into account. Assume at time $t$,  the $i$th device from $\Phi_{\text s}$ has data to transmit. Then, the operation control problem could be written as follows:
 \begin{align}
 \max_{p_i, c_i, h_i, m_i}& F(\text{Rel}_i, \text{EE}_i)\nonumber\\
 \text{s.t: } & p_i\in \mathbb P, c_i\in \mathbb C, h_i\in \mathbb H, m_i\in \mathbb M,\label{op1}
 \end{align}
 in which $F(\cdot)$ represents the objective function in terms of reliability (Rel) and energy-efficiency (EE) of communications. Regarding different QoS requirements of different IoT applications, definition of $F(\cdot)$ may differ from one IoT application to the others. Here, we focus on a weighted sum of objectives, i.e. $F(\text{Rel}_i, \text{EE}_i)=(1-\beta) \text{Rel}_i+\beta \text{EE}_i$, where $0\le \beta \le 1$ offers a tradeoff between reliability and energy efficiency.  Also,  $p_i$, $c_i$, $h_i$, and $m_i$ represent the selected transmit power,  code\footnote{The transmit code also determines the data rate \cite{scal}.}, sub-channel, and number of transmitted replicas per packet\footnote{In the NB-IoT and SigFox, several replicas are transmitted per data packet for range extension and resilience against interference, respectively  \cite{mag_all}.}. Furthermore, $\mathbb X$ represents the set of available values for $X_i$.   In the sequel, we aim at solving this optimization problem.

\section{Self Organization as a Solution}
A centralized solution to the problem in \eqref{op1} is very complex\footnote{In section \ref{lor}, we  investigate this optimization problem analytically to get insight on complexity order of the centralized solution.}, and not applicable to low-power IoT devices which require less frequent signaling with the access network in order to save energy. Thus, instead of solving the problem in a centralized manner, we leverage \textit{distributed online learning}. 
In online learning, each  device, also called hereafter  agent, aims at maximizing its objective function $F(\text{Rel}_i, \text{EE}_i)$ by choosing the best  actions $A_i=\{p_i, c_i, h_i, m_i\}\in \mathbb A$, given the rewards (ACK/NACK) of its previous actions.  After choosing the  action at time $t$, i.e. $A_i(t)$,  agent receives the reward, denoted by $\xi(t)\in\{1,0\}$, where 1 and 0 represent acknowledgment and no acknowledgment respectively. This type of learning is commonly described as {\textit{multi-agent multi-arm bandit (MAB)}} in the literature \cite{sadegh}.  
In MAB learning,  an agent  chooses one of the $K$ arms at each time to play, and receives a reward afterwards. The agent's aim is to maximize its self-accumulative return or equivalently, minimize its self-accumulative regret\footnote{Regret indicates difference between reward of a non-optimal and the optimal action.}. The MAB problem offers a tradeoff between  exploration and exploitation, where the former indicates decision epochs in which agent tries different actions even if their previously observed rewards are less than the others, and the latter indicates decision epochs at which agent acts greedy based on the previous rewards. Due to its  widespread applications in gambling, robotics, and etc., MAB learning has been well investigated in literature, and efficient solutions have been proposed to minimize agent's regret. 
In the sequel, we investigate solutions to the IoT-device's operation control problem in environments  dealing with stochastic  interference over the data channels and no interference over the feedback channel (modeled by stochastic   MAB), as well as environments  dealing with stochastic  interference over the data channels and adversary interference over the feedback channel (modeled by adversary   MAB).
\subsection{Learning for Stochastic MAB}
For stochastic  MAB, the MAB in which  each arm's rewards are drawn from a probability density function (PDF), upper confidence bound (UCB) index policies perform close to optimally \cite{sadegh}. The aim of UCB index policies is to select the arm with  the largest upper confidence bound for the expected return. Then, each agent maintains an index function for each arm, which is a function of the past rewards of this arm, and in each decision epoch, selects the arm with the maximal index.  Among UCB algorithms, the $\text{UCB}_1$ algorithm \cite{aur}, attains a regret growing at $O(\log n)$ in the stochastic MAB, where $n$ is the number of rounds \cite{stosdve}.
\subsection{Learning for Non-stochastic MAB}
Having insights on optimized learning in stochastic MAB, we can investigate learning in non-stochastic MABs, where arms' rewards are not drawn from a specific PDF. An example of non-stochastic MAB is the adversarial environment, in which, an adversary can interrupt the rewards (the feedback messages). Furthermore, in IoT solutions over the unlicensed spectrum, the feedback channel is also affected by the interference from coexisting technologies, and hence, the rewards might be interrupted.   Efficient learning approaches for adversary settings have been proposed in literature, among them, the exponential-weight algorithm for exploration and exploitation (EXP3) is a promising approach \cite{og}.  On each decision epoch $t$, EXP3 chooses an action, out of $\mathbb A$, according to a set of   respective distributions, i.e.  $A_i(t) \sim \{{\text p}_1(t), \cdots, {\text p}_{|\mathbb A|}(t)\}$. EXP3 assigns each action a probability mass function based on mixing the estimated cumulative reward and the uniform distribution, where the former  incurs the  exploitation mode, and the latter incurs the exploration mode. EXP3 attains a regret growing as $O(\sqrt{n})$ in the adversarial MAB \cite{stosdve}.  

\begin{algorithm}[t!]
\nl Initialization: $Z_k(1)\text{=}0, T_k(1)\text{=}1, \forall k\in \mathbb A $\; 
\nl \For{$t=1,2,\cdots$}{
- Update index: $b_k(t)=  Z_k(t)+\sqrt{ {\alpha\log(t)}/{T_k(t)}}$\;
- Take action: $\arg \max_{k\in \mathbb A}\hspace{1mm} b_k(t) \to A(t)$\;
- Receive reward: $\xi(t)\in\{0,1\}$\; 
- Update reward:    $Z_k(t\text{+}1)\text{=}Z_{k}(t), \forall k\in \mathbb A$$\setminus$$A(t)$\;
 \hspace{24mm}   $Z_{A(t)}(t\text{+}1)\text{=}Z_{A(t)}(t)\text{+}\hat z(t)$\;
- Update counter: $T_{A(t)}(t\text{+}1)\text{=}T_{A(t)}(t)\text{+}1$\;
\hspace{25mm} $T_{k}(t\text{+}1)\text{=}T_{k}(t),\forall j\in \mathbb A$$\setminus$$A(t)$\;
- \Return $A(t)$\;
}
\caption{Pseudo-code of $\text{UUCB}_1$. }\label{uuc}
\end{algorithm}
 
 \begin{algorithm}[t!]
\nl Initialization: $W_k(1)=1, \forall k\in \mathbb A$\; 
\nl \For{$t=1,2,\cdots$}{
- Define Dist.: ${\text p}_k(t)\text{=} (1\text{-}\rho)\frac{W_k(t)}{\sum\nolimits_{j=1}^{|\mathbb A|}W_j(t)}\text{+}\frac{\rho}{|\mathbb A|}, \forall k\in \mathbb A$\;
 - Take action: $A(t)\sim\{{\text p}_1(t), \cdots, {\text p}_{|\mathbb A|}(t)\}$\;
- Receive reward: $\xi(t)\in\{0,1\}$\; 
 - Update weight:  $W_k(t\text{+}1)\text{=}W_k(t), \forall k\in\mathbb A$$\setminus$$A(t)$\;
\hspace{16mm} $W_{A(t)}(t\text{+}1)\text{=}W_{A(t)}(t)\exp(\frac{\rho \hat\xi(t)}{|\mathbb A| {\text p}_{A(t)}(t)})$\;
  
- \Return $A(t)$\;
}
\caption{Pseudo-code of $\text{UEXP3}$.}\label{uux}
\end{algorithm}

\subsection{Light-weight Learning for Low-power IoT Networks}
Recall the optimization problem in \eqref{op1}, in which the aim   is to maximize the reliability and energy efficiency of devices. As a distributed solution to this problem, here we incorporate both reliability and energy efficiency in the learning process for stochastic and non-stochastic settings. Let's start with the stochastic setting, where at the end of each successful transmission, the  device receives an acknowledgment. Once the acknowledgment is received, the accumulated reward of the respective action is incremented by one in ${\text{UCB}}_1$ algorithm \cite{aur}. Now, denote by $E_i$ and $E_{\text{min}}$, the consumed energy in packet transmission using action $i$, and the minimum consumed energy amongst actions achieved a successful packet transmission respectively.  In our proposed learning solution, we modify the reward achieved by choosing   action $k$ as:
\begin{equation}\label{ru}
\hat\xi(t)=\xi(t)(1\text{-}\beta)+\xi(t)\beta{E_k}/{E_{\text{min}}}, \forall k\in \mathbb A,
\end{equation}
in which $\xi(t)\in \{0,1\}$ represents the acknowledgment,  $\beta$ is a design parameter offering tradeoff between reliability an energy efficiency, and $t$ represents the time index. 
 Based on this updated reward function, we present the updated $\text{UCB}_1$ ($\text{UUCB}_1$) algorithm  in Algorithm \ref{uuc}. Following the same approach, and by updating the reward function in EXP3 \cite{og}, we present the updated EXP3 (UEXP3) algorithm in Algorithm \ref{uux}. In these algorithms, $\alpha$  and $\rho$ are the design parameters, which offer tradeoff between exploration and exploitation in the $\text{UUCB}_1$ and EXP3 respectively. Furthermore, the device index, i.e. $i$ in the underscript, has been dropped. Mapping \eqref{ru} to the objective function in \eqref{op1}, one sees that $F(\text{Rel}_i,\text{EE}_i)$ in \eqref{op1} has been modeled by the modified reward function, i.e. $\hat\xi(t)=F(\text{Rel}_i,\text{EE}_i)$. Furthermore, the first term in \eqref{ru} represents the external regret, while the second term represents the internal regret.
 In the following section, we employ the proposed learning scheme  in operation control of IoT devices connected through LoRa technology, and compare the results against the results of a centralized optimized solution.

\section{Distributed Learning for IoT Operation Control: A LoRa 
Technology Example}\label{lor}

\subsection{Communication Using The LoRa Technology}
LoRa,  the physical layer of LoRaWAN, aims at enabling
low-power low-rate  long-distance communications. 
Communication in LoRa occurs in 3 sub-channels in the public ISM band; each with bandwidth (BW) of 125 KHz.  High resilience to  noise and interference is the key to operate efficiently in the  ISM band. Towards this end, the chirp spread spectrum (CSS) modulation has been  used in LoRa, which  enables signals with different spreading factors (SFs) to be distinguished and received simultaneously, even if they are transmitted at the same time on the same channel. The spreading factors, ranging from 7 to 12, denote the number of chirps used to encode a bit, and hence, determine the data rate: $R(c)=\frac{c\times \text{BW}\times \mu}{2^{c}}, \forall c\in\mathbb C=\{7,\cdots,12\}$, where $\mu$ is the code-rate.  Based on \cite{scal}, the required SNRs for correct detection of signals with spreading factors $\{7,\cdots,12\}$ are $\gamma_{{\text{th}}_{\text N}}=\{$-6,-9,-12,-15,-17.5,-20$\}$, respectively. Then, one sees that by increase in the SF index, data rate decreases and resilience to   noise  increases.  Finally, LoRaWAN supports the following transmit powers for communication: $\{2, 5, 8,
11, 14 \}$ dBm \cite{psc}. 
\subsection{Operation Control in LoRa}
Consider a LoRa gateway in a 2D plane with multitude of devices, distributed according to a Poisson point process (PPP) with density $\lambda$ in $r\le \mathcal R$, where $r$ is the distance to the BS located at the origin and $\mathcal R$ is the boundary of service area for the gateway. The IoT devices aim at data transfer to the gateway on average each $T_{\text{rep}}$ seconds. Recalling the operation control problem in \eqref{op1} the LoRa operation control consists in solving the following problem:
 \begin{align}
 \max_{p_i, c_i, h_i}& F(\text{Rel}_i,\text{EE}_i) \label{op2}\\
 \text{s.t: } & p_i\in \mathbb P=\{2,5,8,11,14 \} \text{dBm},\nonumber\\
  &c_i\in \mathbb C=\{7,8,9,10,11,12\}, h_i\in \mathbb H=\{1,2,3\}.\nonumber
 \end{align}
\subsubsection{Distributed Learning for  Operation Control}
One can directly apply the presented Algorithms \ref{uuc} and \ref{uux} in section III, in order to solve the optimization problem in \eqref{op2}. In this case, the set of actions, i.e. $\mathbb A$, includes 90 pairs of actions, each including a  transmit power,  sub-channel, and spreading factor. 

\subsubsection{Centralized Optimized Operation Control}
In order to  save space, in this version we present the formulation for $|\mathbb H|$$=$$|\mathbb P|$$=$1, i.e. we consider a single-channel single-transmit power level LoRa network in which, we aim at distributing spreading factors amongst devices. Furthermore, we assume interference is only coming from the coexisting LoRa devices. 
Regarding the fact that by increase in the SF index, data rate decreases, probability of collision increases, and resilience to noise increases, nodes located closer to the BS are expected to choose lower-index SFs and vice versa \cite{eql}. Then, the SF allocation problem is equivalent to finding the optimized density of nodes, which are  using different spreading factors in each region of the service area. Let us divide the service area to a set of rings, each with inner and outer radius of $r_{j,1}$ and $r_{j,2}$ respectively, where in each of them, density of nodes which are using each SF is assumed to be constant ($j\in \mathbb J$). By extending the  results in \cite[Section~III.A]{opd}, one can derive\footnote{Details will be presented in the journal version.} the Laplace functional of interference from  devices distributed on the $j$th ring, denoted by $\Phi_{j,c}\subset \Phi$, with transmit power $P_t$, reporting period $T_{\text{rep}}$, packet length $D$, spreading factor $C$, and  transmission time $T(c) =D/R(c)$, as:
 \begin{align}
 \mathcal L_{\Phi_{j,c}}(s)&\text{=} \exp\big( -2\pi   \int\nolimits_{r_{j,1}}^{r_{j,2}}\frac{ \lambda_{j,c} {T_{c}}/{T_{\text{rep}}}}{\frac{1}{s    P_{\text t} G r^{-\delta} } \text{+}1}  { r  d r}\big),
\nonumber
 \end{align}
in which  $Gr^{-\delta}$  is  the pathloss.  Now, the Laplace functional of received interference from all devices using spreading factor $c$ could be written as $\mathcal L_{\Phi_{c}}(s)=\prod_{j\in \mathbb J}  \mathcal L_{\Phi_{j,c}}(s)$. Let $N$  and $I_\Phi$ denote the additive
noise and interference from set $\phi$ of devices at the receiver. Using the above derived  interference model,  probability of success in  packet transmission for a device located at distance $ z$ to the BS, using spreading factor $c\in \mathcal C$, is derived as: 
\begin{align}
\text{p}_{\text{s}}(c,{ z})&\text{=}\text{Pr}({  P_{\text{t}}h Gz^{-\delta}}\ge  \gamma_c N) \text{Pr}({  P_{\text{t}}h Gz^{-\delta}}\ge  \gamma_{{\text{th}}_{\text I}} I_{{\Phi}_c}),   \label{psucs}\\
  &{\text =} \mathcal L_{{{\Phi}_c}}(s) \big|_{s=\frac{\gamma_c }{ P_{\text{t}}  Gz^{-\delta}}}\mathcal L_{N}(s)  \big|_{s=\frac{\gamma_{{\text{th}}_{\text I}} }{ P_{\text{t}}  Gz^{-\delta}}},\nonumber\\
 &\text{=}    \prod\limits_{j\in \mathbb J}  \exp\big( \int\nolimits_{r_{j,1}}^{r_{j,2}}\frac{\text{-}2\pi \lambda_{j,c}  {T_{c}}/{T_{\text{rep}}}}{1\text{+}(\frac{r}{z})^\delta\frac{1}{   {\gamma_{\text{th}}}_{\text I}}}  { r  d r}\big)
\exp\big(\text{-}\frac{\mathcal N\gamma_c (c)z^\delta}{  P_{\text{t}} G} \big),\nonumber\\
&\text{=}   \prod\limits_{j\in \mathbb J} \exp\big(\text{-}\lambda_{j,c}  \frac{T_{ c}}{T_{\text{rep}}} [Q(r_{j,2})\text{-}Q(r_{j,1})]\big)\exp\big(\text{-}\frac{ \mathcal N\gamma_c z^\delta}{  P_{\text{t}} G} \big),\nonumber
\end{align}
where $\gamma_c= \gamma_{{\text{th}}_{\text N}}(c)$, and   for $\delta=4$, we have\footnote{Similar expressions could be found using table of integrals for other pathloss exponent values.}: 
$$Q(x)=\pi {\text{atan}(\frac{1}{\sqrt{ {\gamma_{\text{th}}}_{\text I}}}\frac{x^2}{z^2})}/{\frac{1}{\sqrt{ {\gamma_{\text{th}}}_{\text I} }z^2}}.$$
Now, the optimization problem in \eqref{op2} reduces to:
 \begin{align}
 \max_{\lambda_{j,c}}& \hspace{2mm}\sum\limits_{j\in\mathbb J} \sum\limits_{c\in\mathbb C}\frac{\lambda_{j,c}}{\lambda} \bigg[(1-\beta)\int\nolimits_{r_{j,1}}^{r_{j,2}}\text{p}_{\text{s}}(c,{ z})dz+\beta \frac{T_c}{T_1} \bigg],  \label{op3}\\
 \text{s.t: } & \sum\nolimits_{c\in \mathbb C}\lambda_{j,c}=\lambda.\nonumber
 \end{align}
By solving this problem, the density of usage of each SF is identified as a function of distance to the BS. From \eqref{op3}, one sees that solving the centralized optimization problem, even in the simplified form where we  assumed a 2D PPP distribution   with a single-level transmit power and a single-channel LoRa network without external interferer, is highly complicated. In the following, we compare performance of the centralized optimized solution against the distributed learning approach.
\subsubsection{Comparison of Solutions}
\begin{figure}[t!]
        \centering
                \includegraphics[width=3in]{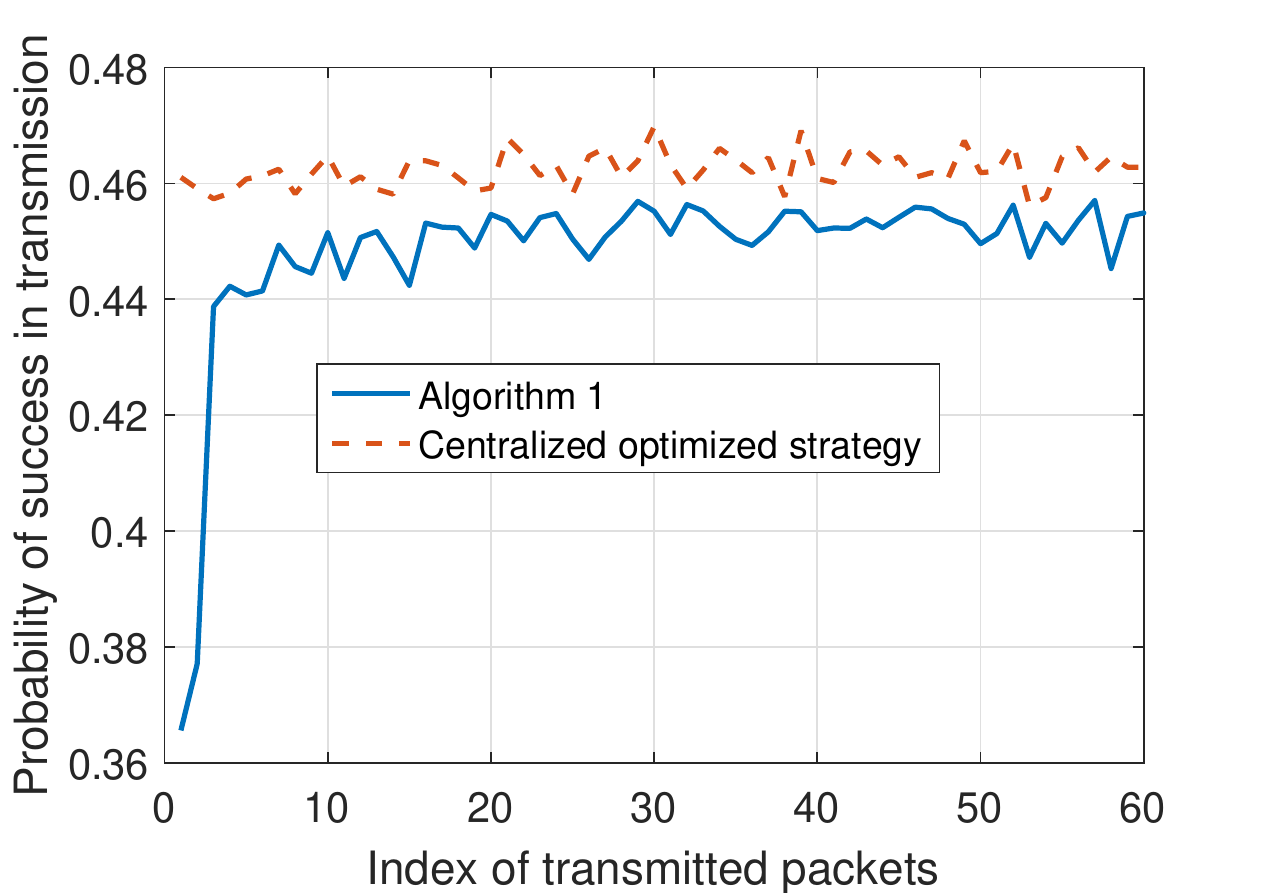}
     \caption{Learning for SF selection control: comparison of Alg1 and the centralized optimized solution. }
                     \label{suc-opt}
 \end{figure}

Fig. \ref{suc-opt} compares  probability of success in transmission  for Algorithm \ref{uuc} and the result of centralized optimized strategy for the case: $\mathbb C=\{7,10\}$, ${\gamma_{\text{\text{th}}}}_\text{N}$$=$$\{$-6,-15$\}$ dB, $P_t$=14 dBm, $N_d=1000$, $T_{\text{rep}}=200$ sec, and $D$= 100 bytes. Other parameters could be found in Table \ref{part}. The $x$-axis represents index of transmitted packets. Here, each node  decides to send data over SF 7 or 10, i.e. $|\mathbb A|=2$. One sees that after a few number of transmissions, the learning's results become close to the centralized solution. These results also promise a valuable improvement in the battery lifetime due to the fact that without need for listening to the BS and signaling, we have configured communication parameters of IoT devices in a distributed form. 
Detailed energy and reliability performance evaluations are presented in the following section.

\begin{table}[t!]
 \centering \caption{Parameters for performance analysis  \cite{scal}.}\label{part}
\begin{tabular}{ p{4 cm}p{4 cm}}\\
\toprule[0.5mm]
Parameters &Values\\
\midrule[0.3mm]
Service area & Circle of radius 2 Km\\
Aggregated packet arrival rate: $\sigma= N_d /T_{\text{rep}}$& 12.5 (Sc1), 2.5 (Sc2,3) per seconds\\
Packet length: $D$ & 100 bytes (Sc1), 20 bytes (Sc2,3)\\ 
Number of sub-channels& 1 (Sc1,2), 3 (Sc3)\\
Bandwidth: $\mathcal W$&125 KHz\\
Code rate: $\mu$ & 4/5\\
Threshold SNR: $\gamma_{\text{\text{th}}_N}$
& $\{$-6,-9,-12,-15,-17.5,-20$\}$ dB\\
Threshold SIR: $\gamma_{\text{\text{th}}_I}$& 6 dB\\
Power consumption: $P_t$, $P_c$, $\eta$& \{8,14\} dBm, 10 dBm, 2  \\
Learning parameters: $\alpha$, $\beta$, $\rho$ & 0.1, 0.5, 0.4\\
\bottomrule[0.5mm]
\end{tabular}
\end{table}

\section{Performance Evaluation}
In this section, we present the simulation results in the context of  operation control for the LoRa technology. We assume a massive number of LoRa nodes have been distributed according to a PPP in a cell of radius 2 Km. Our aim is to distribute 6 spreading factors and two transmit power levels amongst them.  The simulation parameters have been presented in  Table \ref{part}.  In this table, Sc1, Sc2, and Sc3 refer to three different scenarios in which, simulations have been carried out. In the following figures, Alg1 refers to the Algorithm \ref{uuc}, Alg2 refers to the Algorithm \ref{uux}, EqLoad refers to the centralized algorithm proposed in \cite{eql}, in which number of devices using a SF is  proportional to its data rate, and RandSel refers to the algorithm in which SFs are selected randomly. The aforementioned schemes differ in the way they choose the SF, while all of them choose $P_t= 14$ dBm as the transmit power. In contrast, Alg1(PC) refers to the Algorithm \ref{uuc} in which, devices have freedom to choose their spreading factors and transmit powers out of $\mathbb C$ and $\mathbb P$, respectively.

\begin{figure}[t!]
        \centering
         \begin{subfigure}[t]{0.47\textwidth}
         \centering
                \includegraphics[width=3in]{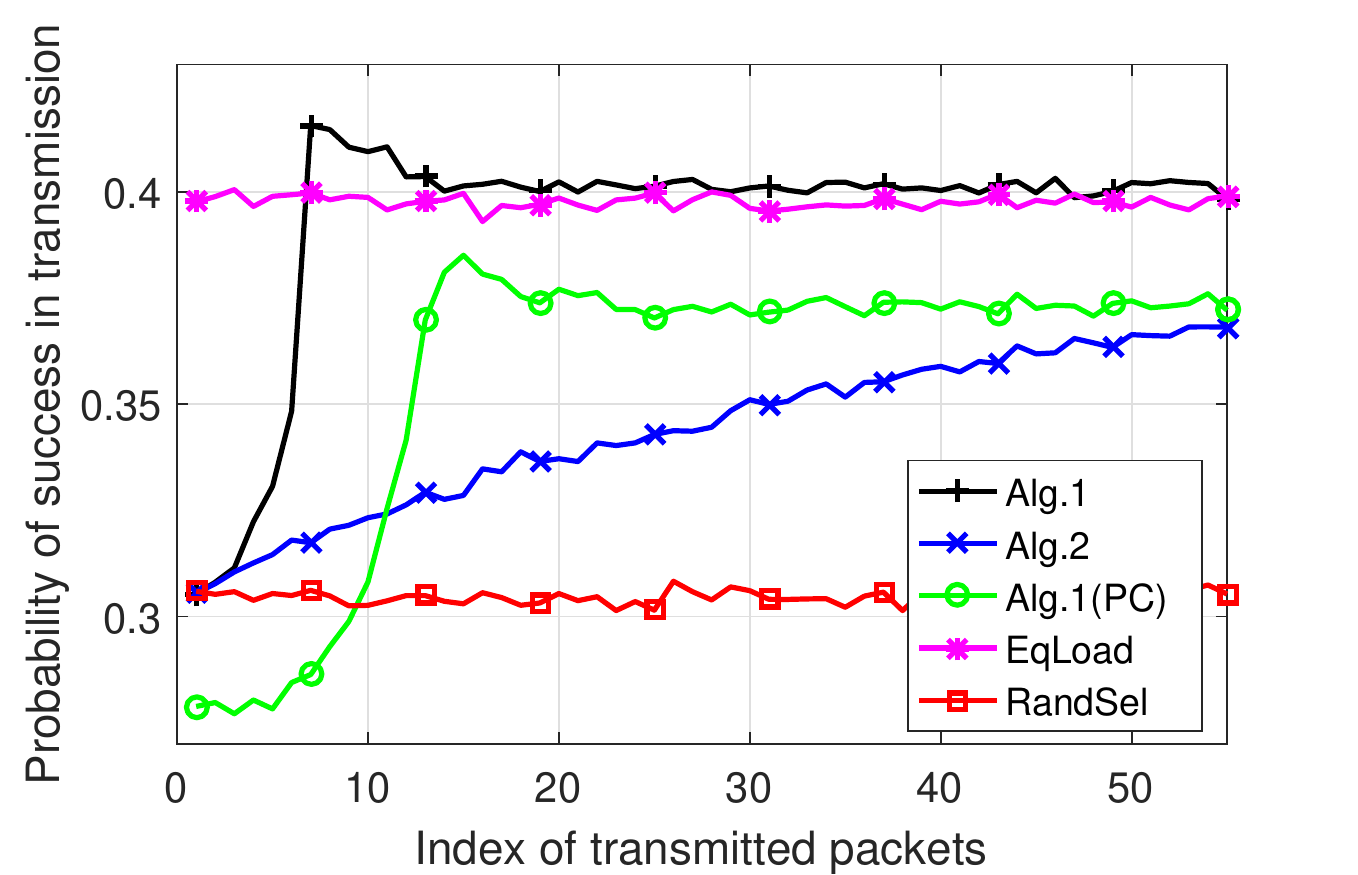}
                \label{suc-nint}
\end{subfigure}\\
         \begin{subfigure}[t]{0.47\textwidth}
        \centering
                \includegraphics[width=3in]{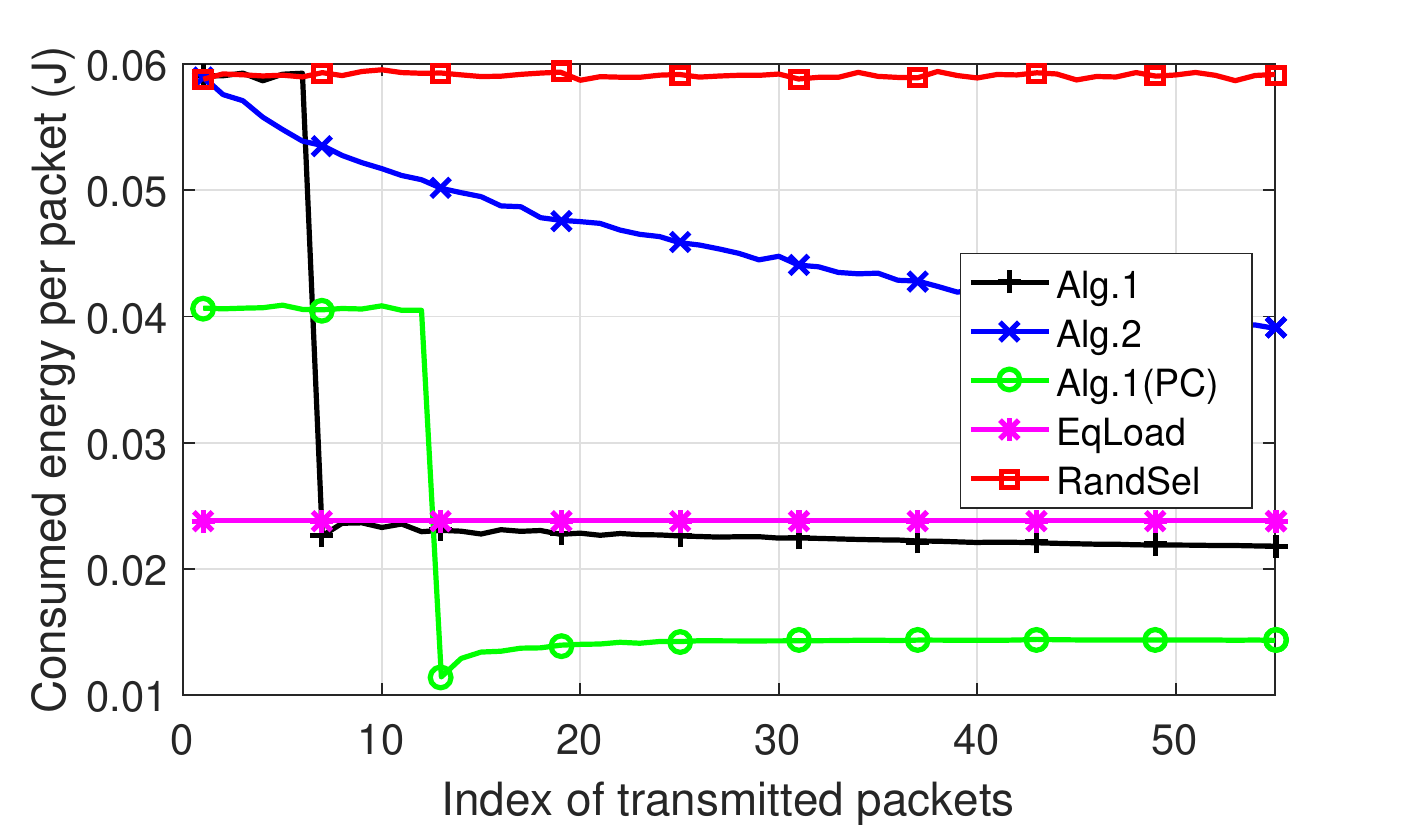}
                \label{ene-nint}
\end{subfigure}
     \caption{Learning for power and SF selection control without external interference (Sc1).}
                \label{no_int}
 \end{figure}

\begin{figure}[t!]
        \centering
         \begin{subfigure}[t]{0.23\textwidth}
        \centering
                \includegraphics[width=2in]{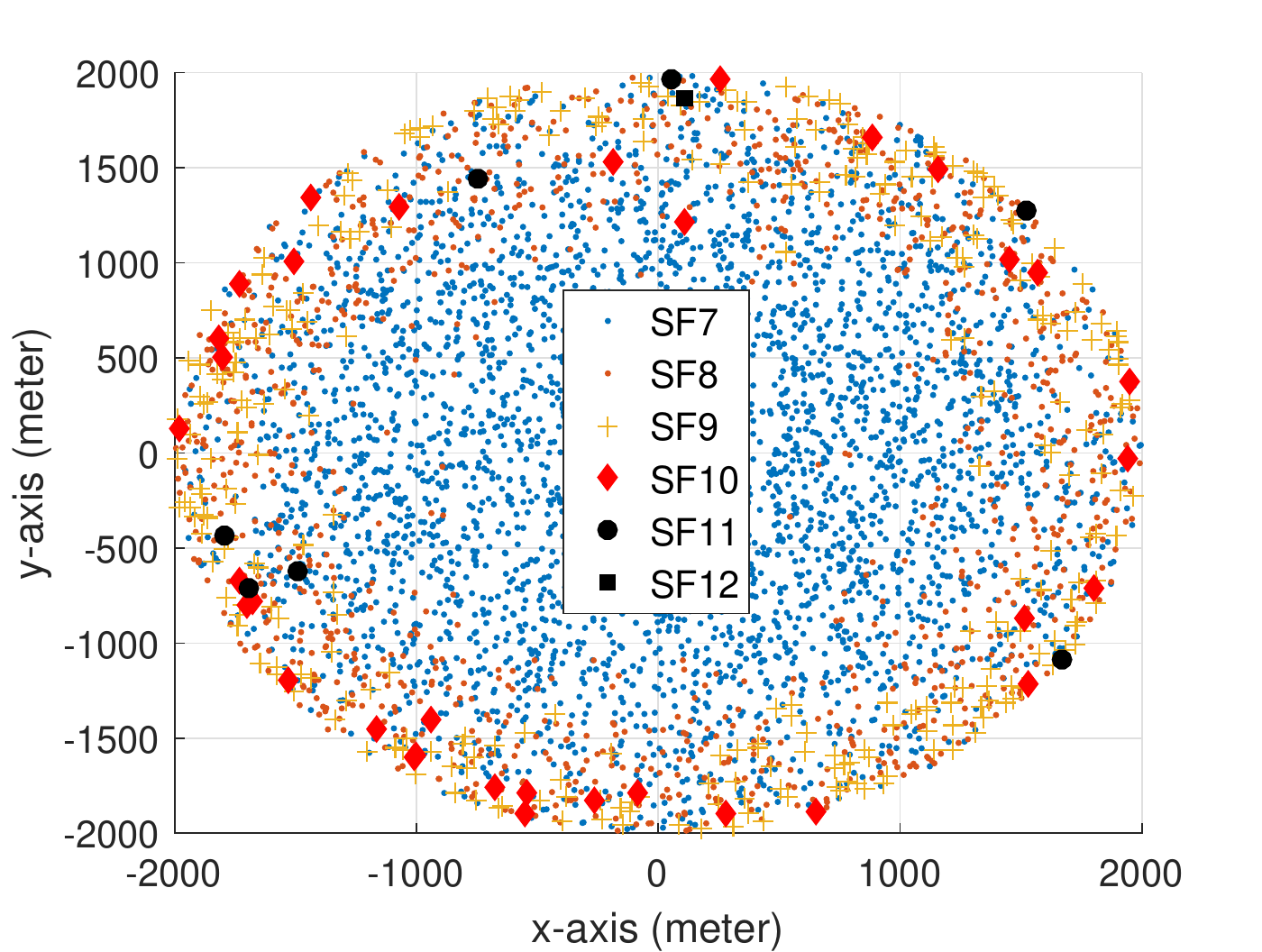}
                \label{cons1}
\end{subfigure}~
 \begin{subfigure}[t]{0.23\textwidth}
        \centering
                \includegraphics[width=2in]{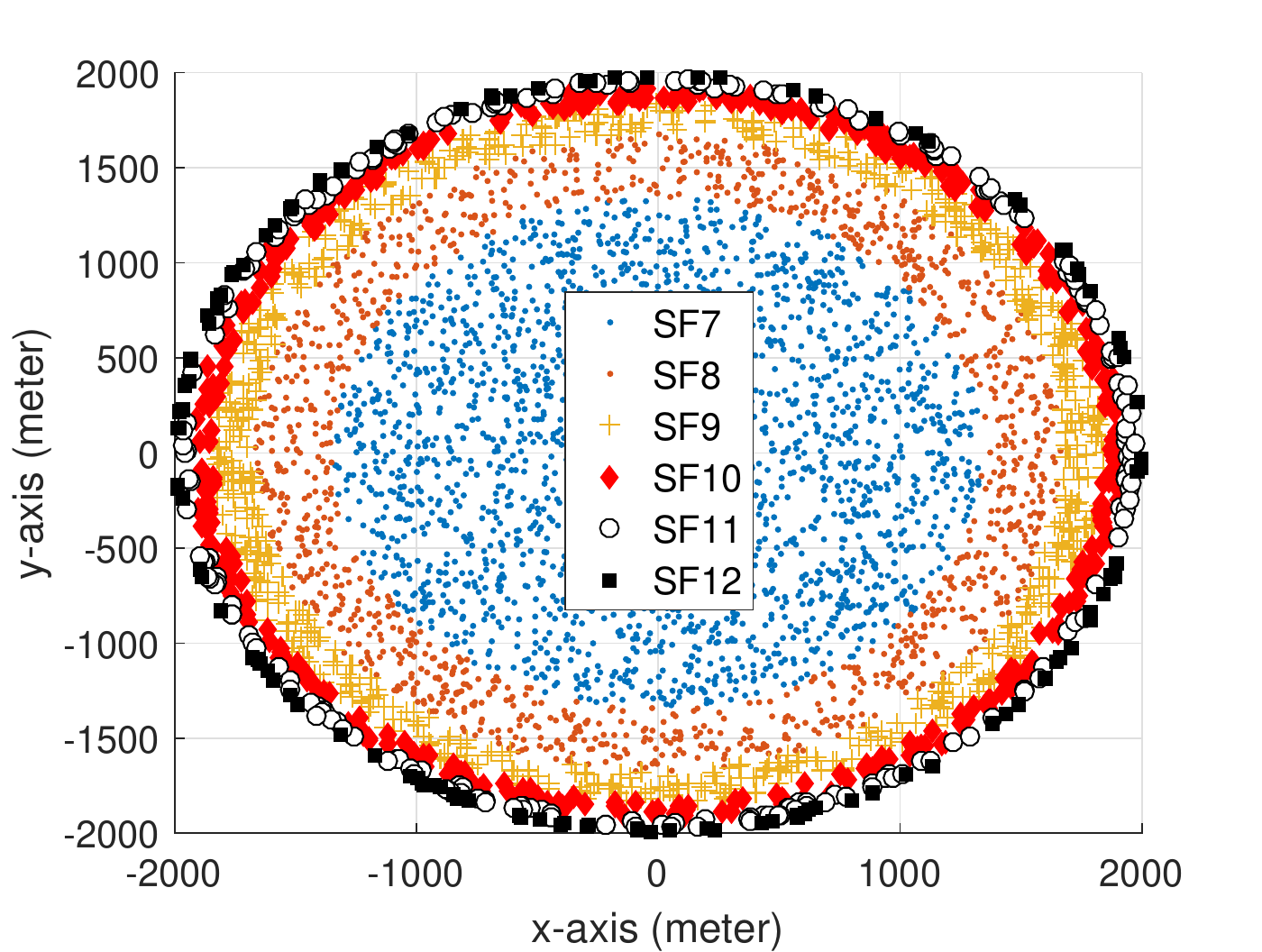}
                \label{cons5}
\end{subfigure}
     \caption{The constellations of selected SFs for Alg1 (left) and EqLoad Algorithm (right).}
                \label{cons}
 \end{figure}

Fig. \ref{no_int} represents reliability and energy efficiency performance evaluations of different schemes versus index of transmitted packets in a scenario in which there is no external interference. One sees that Alg1 is converging   after a few transmissions and is able to have the superior performance in success probability and energy efficiency, even in comparison with the centralized solution proposed in \cite{eql}.   Fig. \ref{cons} represents how different SFs have been allocated to devices in different regions of the cell  by following Alg1 (left) and the EqLoad scheme (right) \cite{eql}. 

Fig. \ref{int} represents the reliability and energy consumption performance evaluations in the same setup as for Fig. \ref{no_int}, with the only difference that here  external stochastic interference has been considered. The probability of occurring interference on each SF differs from the others. One sees that huge increase in probability of success and decrease in energy consumption could be achieved by using the proposed learning approaches. Furthermore, this figure presents the  tradeoff between reliability and energy efficiency, which can be controlled by the design parameter, $\beta$. In this regard, one sees that Alg1(PC) achieves an acceptable success probability in data transmission with an ultra-low energy consumption profile by using $\beta=0.5$ (the solid green line). On the other hand, by decreasing $\beta$ to 0.01, one sees that probability of success has been   significantly improved while energy consumption has been increased in comparison with the previous state. One must note that the energy consumption results in this section represent the energy consumption per packet transmission trial, and hence, the ultimate decrease in energy consumption of devices using our learning approaches will be much higher due to the following facts. (i) The probability of success achieved using  the learning approaches is higher than the other schemes, which on the other hand implies  less number of required retransmissions. (ii) The learning  approaches do not need frequent listening to the BS and receiving control data from them, which on the other hand implies significant reduction in energy consumption for coordination.
 
\begin{figure}[t!]
        \centering
         \begin{subfigure}[t]{0.47\textwidth}
         \centering
                \includegraphics[width=3in]{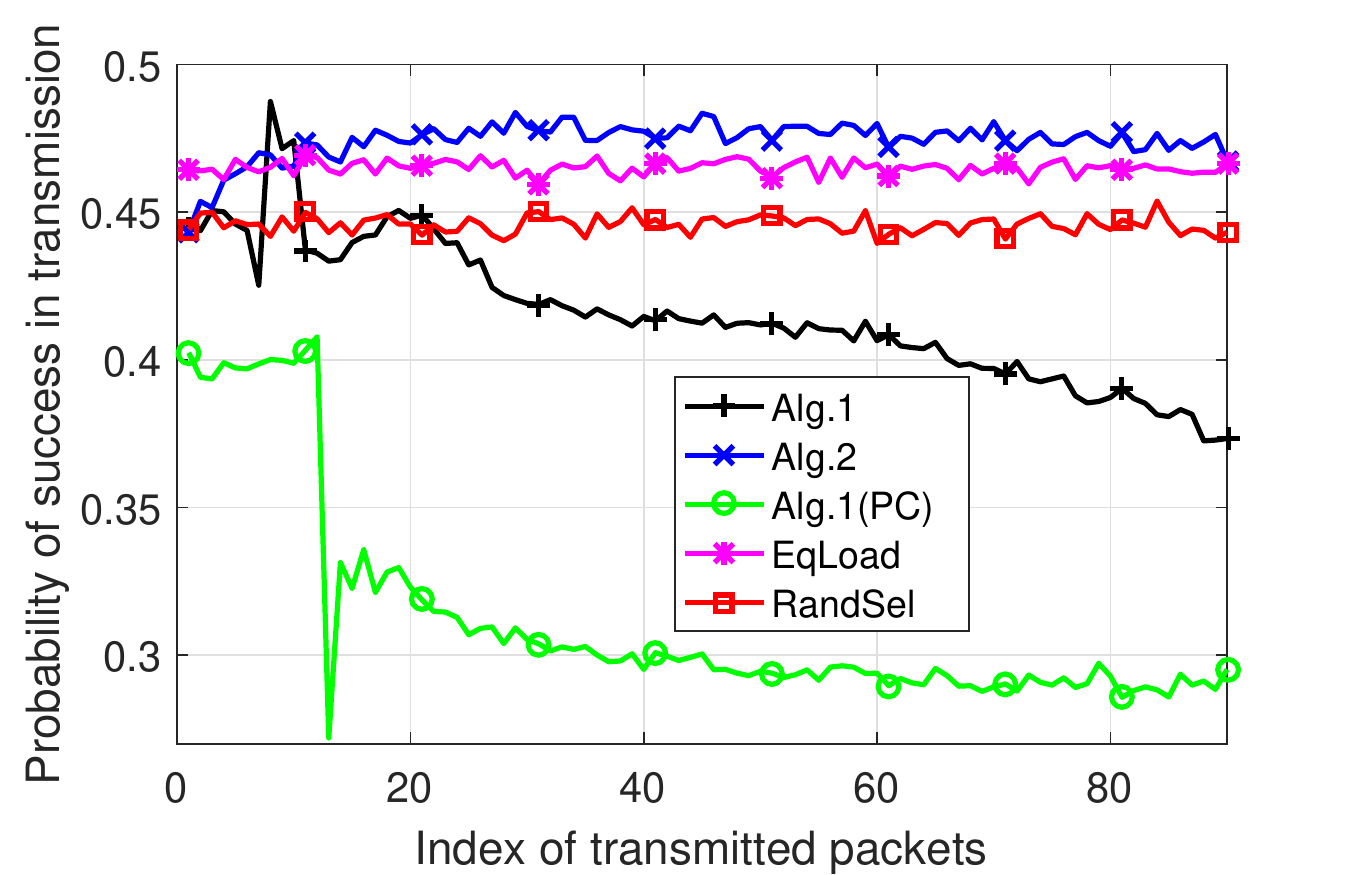}
                \label{suc-adv}
\end{subfigure}\\
         \begin{subfigure}[t]{0.47\textwidth}
        \centering
                \includegraphics[width=3in]{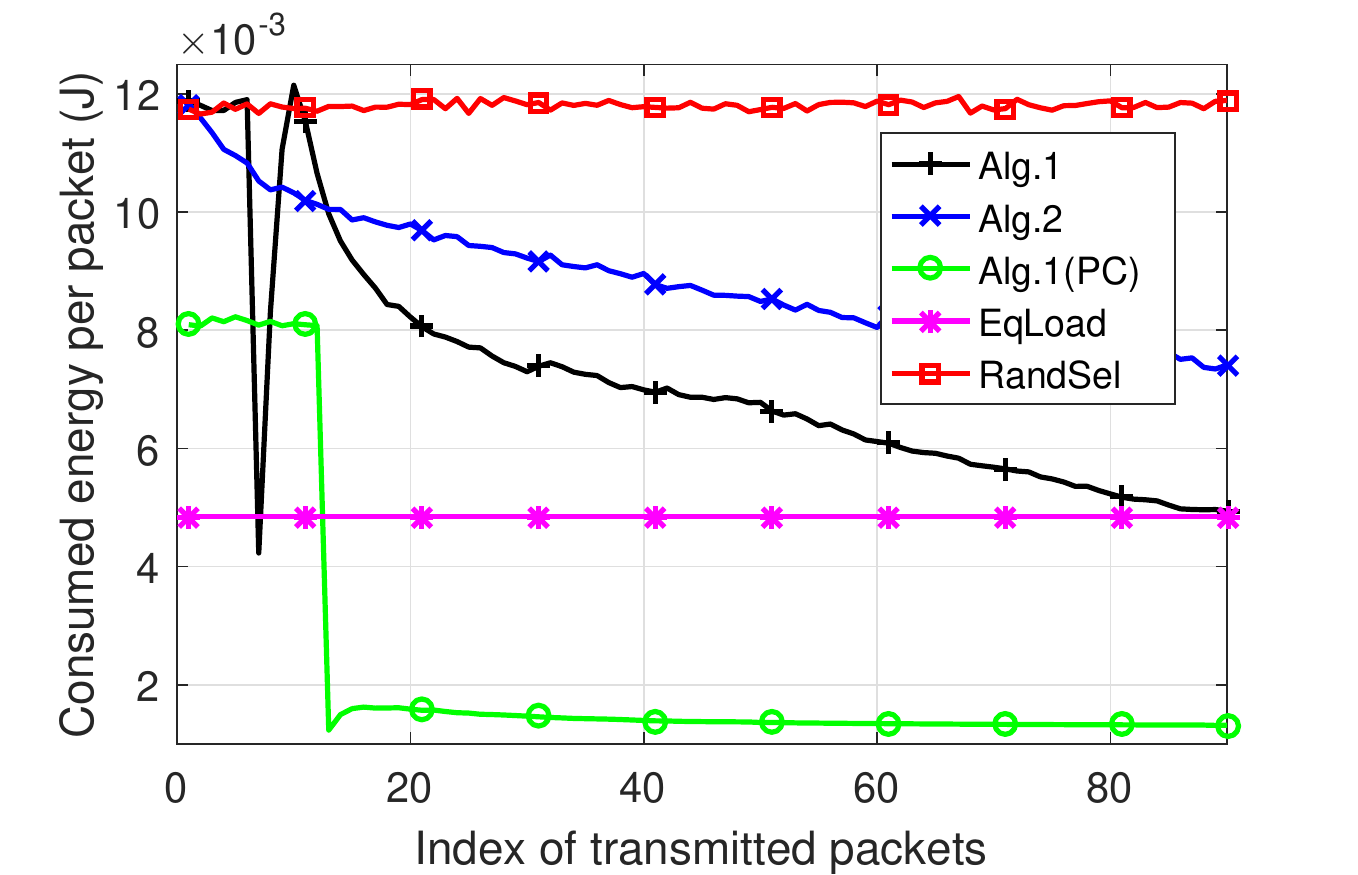}
                \label{ene-adv}
\end{subfigure}\\
     \caption{Learning for power and SF selection control in an adversarial setting with stochastic external interference (Sc2). }
                \label{adv}
 \end{figure}

 Fig. \ref{adv} represents the energy consumption and reliability results in an adversarial setting where, the adversary affects 50\% of the feedback messages sent by the BS. In other words,  with  probability of 0.5,  an ACK message is substituted by a NACK message and vice versa. One sees that in adversary environments, Alg2 outperforms the others in reliability and energy consumption performance measures. One must note that the energy consumption per packet transmission for Alg1 is lower than Alg2. However, due to the increased number of required retransmissions in Alg1, the total energy consumption of Alg1 will be higher than Alg2.  

\begin{figure}[t!]
        \centering
         \begin{subfigure}[t]{0.47\textwidth}
         \centering
                \includegraphics[width=3in]{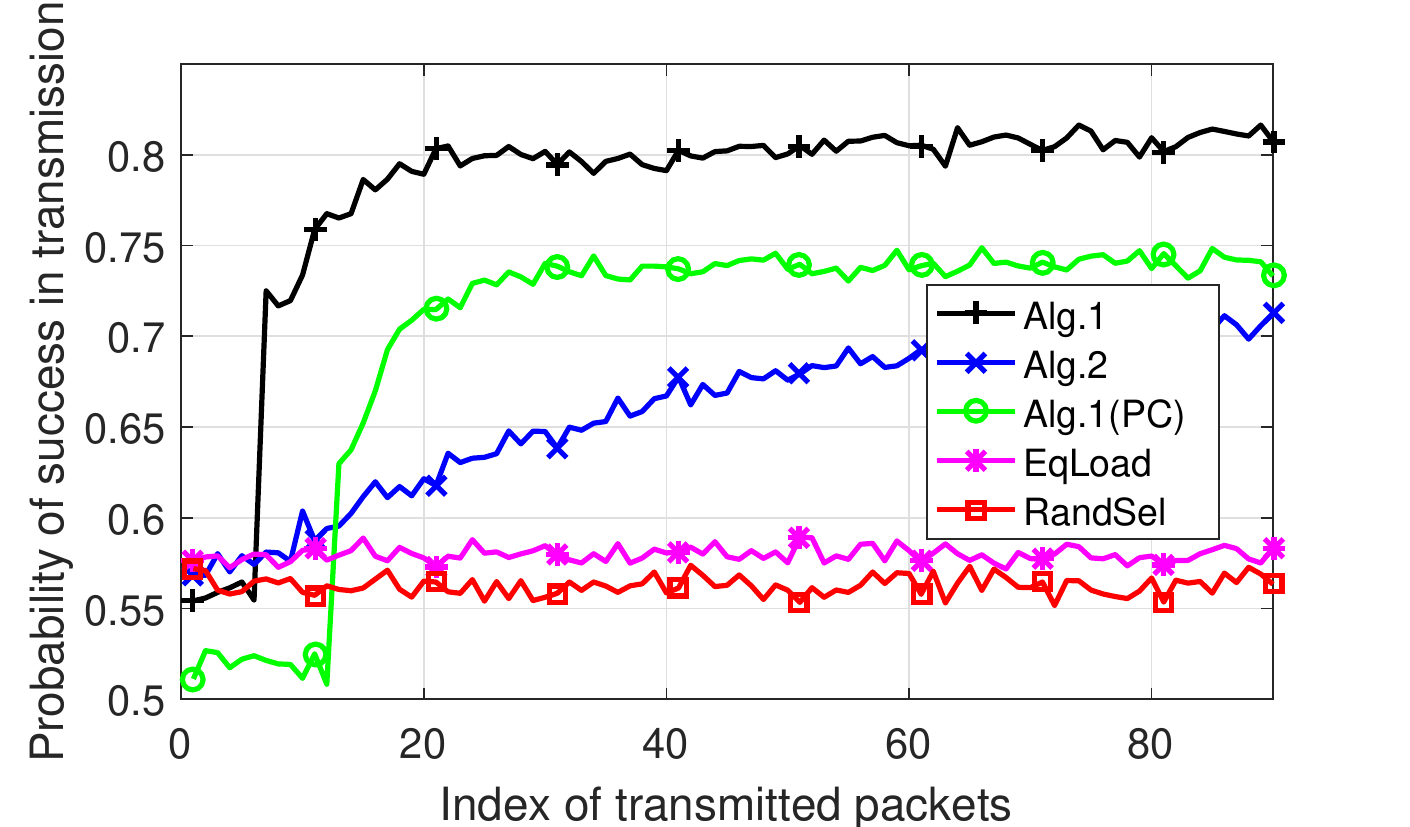}
                \label{suc-chan}
\end{subfigure}\\
         \begin{subfigure}[t]{0.47\textwidth}
        \centering
                \includegraphics[width=3in]{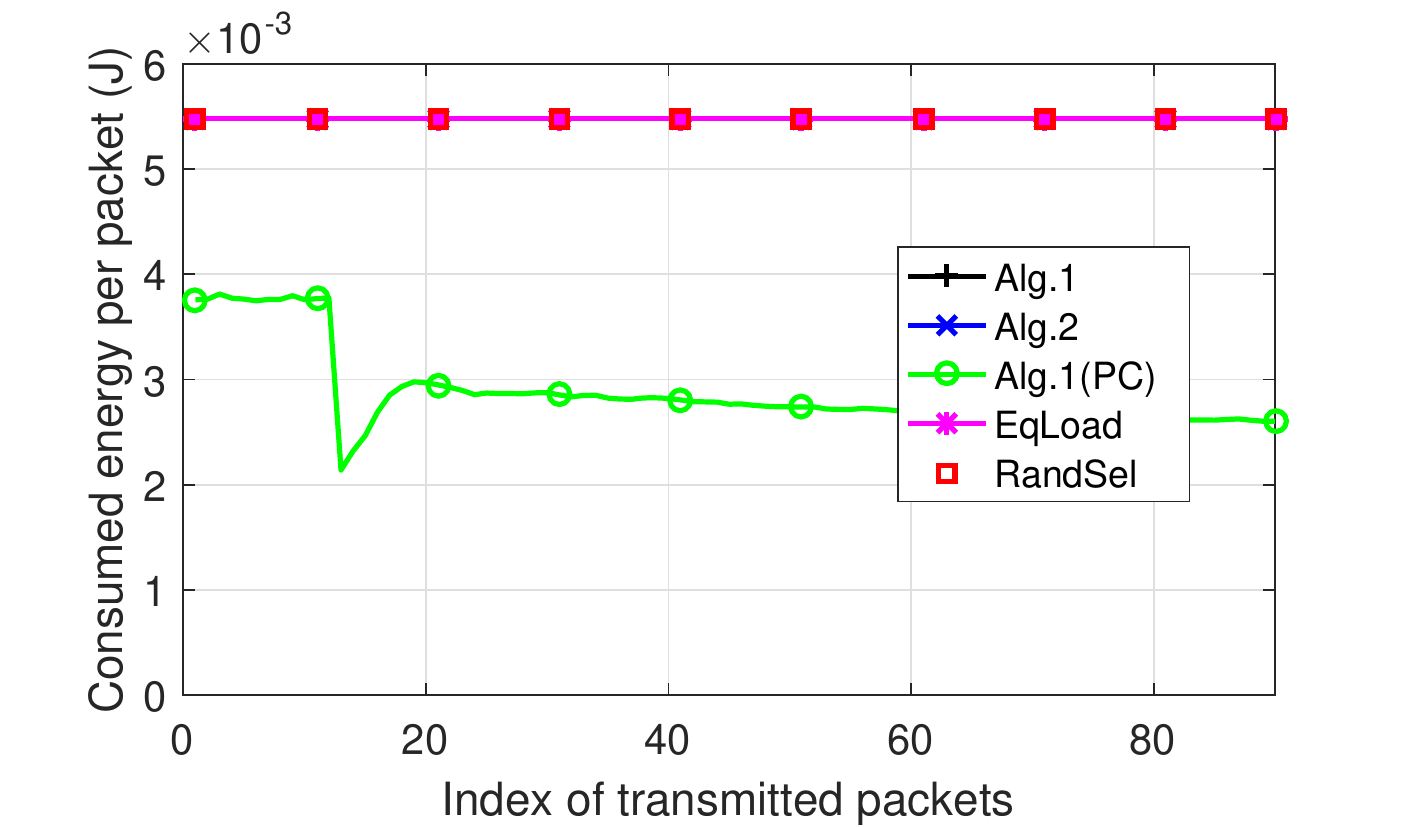}
                \label{ene-chan}
\end{subfigure}\\
     \caption{Learning for power and sub-channel selection control with stochastic external interference (Sc3). All devices are using SF 9.}
                \label{chan}
 \end{figure}

Fig. \ref{chan} represents the energy efficiency and reliability results for the problem in which devices learn to send data over sub-channels  suffering from different levels of stochastic interference.  One sees that the Alg1(PC) scheme is able to achieve more than 50\% decrease in energy consumption per data transfer, while  increasing the probability of success  by 30\% in comparison with the benchmark schemes.   

 \begin{figure}[t!]
        \centering
         \begin{subfigure}[t]{0.47\textwidth}
         \centering
                \includegraphics[width=3in]{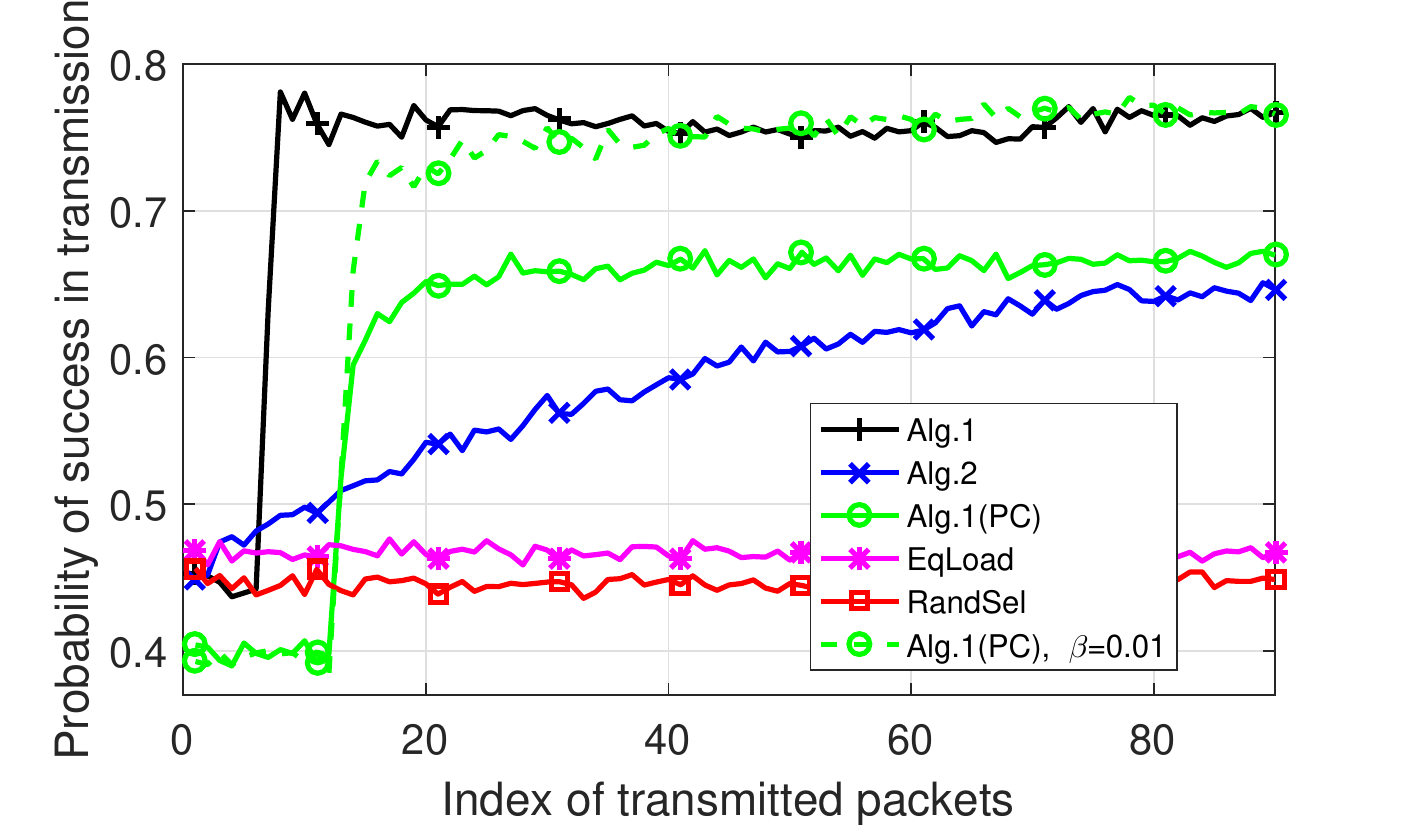}
                \label{suc-int}
\end{subfigure}\\
         \begin{subfigure}[t]{0.47\textwidth}
        \centering
                \includegraphics[width=3in]{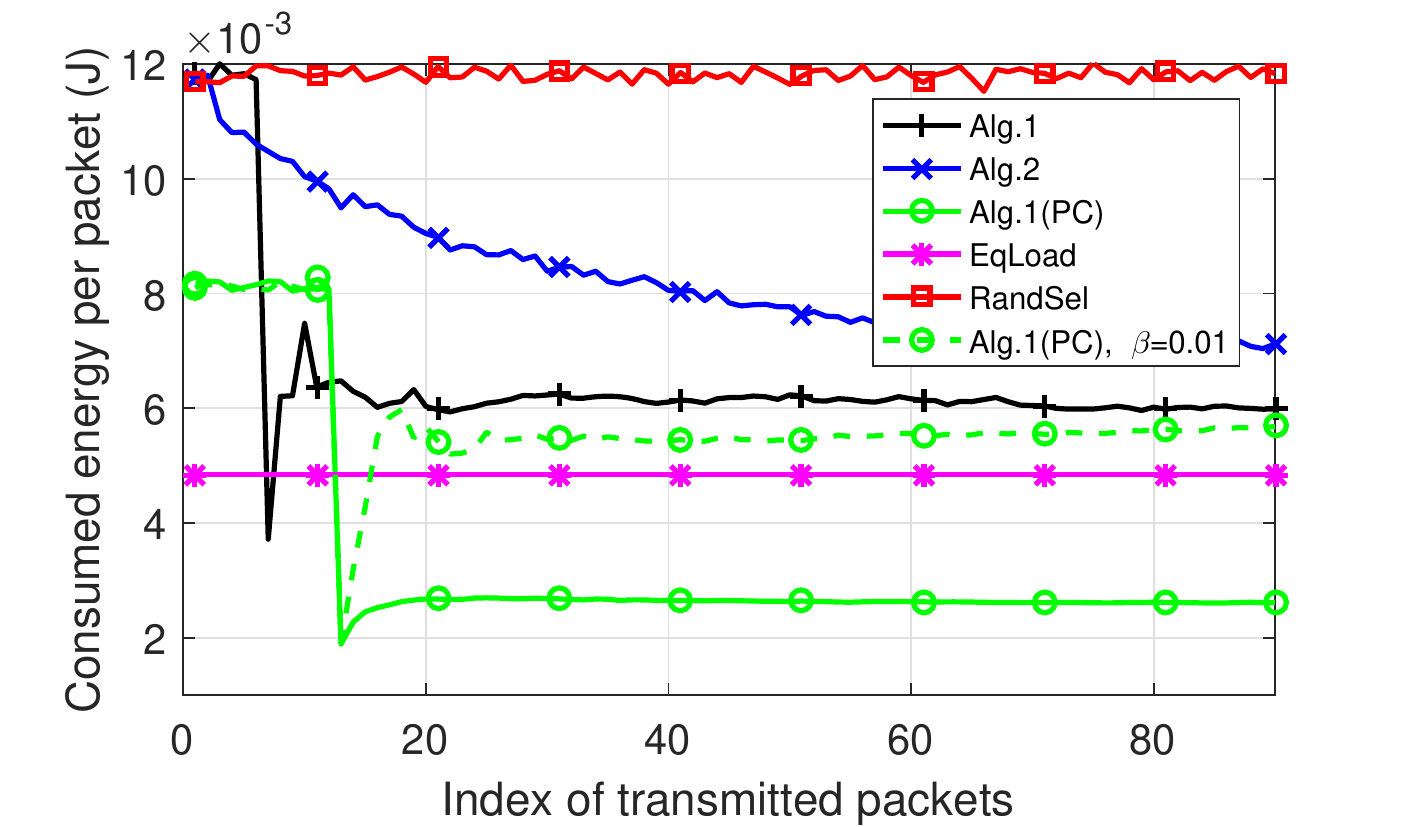}
                \label{ene-int}
\end{subfigure}\\
     \caption{Learning for power and SF selection control with stochastic external interference (Sc2). }
                \label{int}
 \end{figure}

 \section{Conclusion}
Distributed learning for IoT communications with reduced network intervention has been investigated. Reducing signaling between IoT devices and the access network results in decreasing energy consumption  per data transfer for IoT devices as well as  decreasing control over   radio resource usage for the access network. In order to benefit from the former, and suffer as low as possible from the latter, here we have presented a light-weight distributed  learning approach, to be implemented in the device-side. The presented approach leverages external and internal regrets, for minimizing energy consumption and collision probability, respectively, in data transmission over shared wireless channels. The proposed approach has been subsequently employed in operation control of IoT devices using LoRa technology, where analytical as well as simulation results have been derived to characterize performance of the proposed distributed learning approach versus the centralized optimized approach.  The analytical and simulation results represent significant improvement in probability of success in data transmission  as well as battery lifetime of devices by utilizing the proposed learning approach, even in adversarial setups. These results confirm that equipping  IoT  devices with  lightweight learning enables them to adapt themselves effectively to the environment, and hence, makes communications more reliable and durable with reduced network intervention.
 
\ifCLASSOPTIONcaptionsoff
  \newpage
\fi

\bibliographystyle{IEEEtran}
\bibliography{mab_v1}

\begin{thebibliography}{10}
\providecommand{\url}[1]{#1}
\csname url@samestyle\endcsname
\providecommand{\newblock}{\relax}
\providecommand{\bibinfo}[2]{#2}
\providecommand{\BIBentrySTDinterwordspacing}{\spaceskip=0pt\relax}
\providecommand{\BIBentryALTinterwordstretchfactor}{4}
\providecommand{\BIBentryALTinterwordspacing}{\spaceskip=\fontdimen2\font plus
\BIBentryALTinterwordstretchfactor\fontdimen3\font minus
  \fontdimen4\font\relax}
\providecommand{\BIBforeignlanguage}[2]{{%
\expandafter\ifx\csname l@#1\endcsname\relax
\typeout{** WARNING: IEEEtran.bst: No hyphenation pattern has been}%
\typeout{** loaded for the language `#1'. Using the pattern for}%
\typeout{** the default language instead.}%
\else
\language=\csname l@#1\endcsname
\fi
#2}}
\providecommand{\BIBdecl}{\relax}
\BIBdecl

\bibitem{pet}
P.~Popovski, ``Will communications theory finally make itself redundant?''
  \emph{{IEEE ComSoc Technology News}}, 2017.

\bibitem{5g_iot}
C.~Mavromoustakis, G.~Mastorakis, and J.~M. Batalla, \emph{{Internet of Things
  (IoT) in {5G} mobile technologies}}.\hskip 1em plus 0.5em minus 0.4em\relax
  Springer, 2016.

\bibitem{datad}
M.~Kulin, C.~Fortuna, E.~De~Poorter, D.~Deschrijver, and I.~Moerman,
  ``Data-driven design of intelligent wireless networks: An overview and
  tutorial,'' \emph{Sensors}, vol.~16, no.~6, p. 790, June 2016.

\bibitem{sysreq}
{3GPP TS 22.368 V13.1.0}, ``Service requirements for machine-type
  communications,'' Tech. Rep., 2014.

\bibitem{nok1}
{Nokia Networks}, ``{LTE-M} -- optimizing {LTE} for the {Internet} of things,''
  Tech. Rep., 2015.

\bibitem{mag_all}
W.~Yang, M.~Wang, J.~Zhang, J.~Zou, M.~Hua, T.~Xia, and X.~You, ``Narrowband
  wireless access for low-power massive internet of things: A bandwidth
  perspective,'' \emph{IEEE Wireless Commun.}, vol.~24, no.~3, pp. 138--145,
  2017.

\bibitem{lif_com}
{\'E}.~Morin, M.~Maman, R.~Guizzetti, and A.~Duda, ``Comparison of the device
  lifetime in wireless networks for the internet of things,'' \emph{IEEE
  Access}, vol.~5, pp. 7097--7114, 2017.

\bibitem{rsma}
{R1-163510 }, ``{Candidate NR Multiple Access Schemes },'' Tech. Rep., April
  2016, 3GPP TSG RAN WG1 Meeting 84, Busan, Korea.

\bibitem{int2}
M.~Lauridsen \emph{et~al.}, ``Interference measurements in the european 868
  {MHz ISM band with focus on LoRa and SigFox},'' in \emph{IEEE WCNC}, 2017,
  pp. 1--6.

\bibitem{meysam}
M.~Masoudi, A.~Azari, E.~A. Yavuz, and C.~Cavdar, ``{Grant-free Radio Access
  IoT Networks: Scalability Analysis in Coexistence Scenarios},'' in \emph{IEEE
  ICC}, 2018.

\bibitem{eql}
F.~Cuomo \emph{et~al.}, ``{EXPLoRa: EXtending the Performance of LoRa by
  suitable spreading factor allocations},'' in \emph{IEEE 13th International
  Conference on Wireless and Mobile Computing, Networking and Communications
  (WiMob)}, 2017, pp. 1--5.

\bibitem{learn5g}
C.~Jiang, H.~Zhang, Y.~Ren, Z.~Han, K.~C. Chen, and L.~Hanzo, ``Machine
  learning paradigms for next-generation wireless networks,'' \emph{IEEE
  Wireless Communications}, vol.~24, no.~2, pp. 98--105, April 2017.

\bibitem{enb}
Y.~J. Liu, S.~M. Cheng, and Y.~L. Hsueh, ``{eNB} selection for machine type
  communications using reinforcement learning based markov decision process,''
  \emph{IEEE Transactions on Vehicular Technology}, vol.~66, no.~12, pp.
  11\,330--11\,338, Dec. 2017.

\bibitem{adhoc}
A.~Azari, ``Energy-efficient scheduling and grouping for machine-type
  communications over cellular networks,'' \emph{Ad Hoc Networks}, vol.~43, pp.
  16 -- 29, 2016.

\bibitem{mab7}
R.~Bonnefoi, L.~Besson, C.~Moy, E.~Kaufmann, and J.~Palicot, ``Multi-armed
  bandit learning in {IoT} networks: Learning helps even in non-stationary
  settings,'' in \emph{CROWNCOM}, 2017.

\bibitem{seciot}
L.~Xiao, X.~Wan, X.~Lu, Y.~Zhang, and D.~Wu, ``{IoT} security techniques based
  on machine learning,'' \emph{arXiv preprint arXiv:1801.06275}, 2018.

\bibitem{scal}
O.~Georgiou and U.~Raza, ``{Low power wide area network analysis: Can LoRa
  scale?}'' \emph{IEEE Wireless Communications Letters}, vol.~6, no.~2, pp.
  162--165, 2017.

\bibitem{sadegh}
M.~S. Talebi Mazraeh~Shahi, ``Minimizing regret in combinatorial bandits and
  reinforcement learning,'' Ph.D. dissertation, KTH Royal Institute of
  Technology, 2017.

\bibitem{aur}
P.~Auer, N.~Cesa-Bianchi, and P.~Fischer, ``Finite-time analysis of the
  multiarmed bandit problem,'' \emph{Machine learning}, vol.~47, no. 2-3, pp.
  235--256, 2002.

\bibitem{stosdve}
S.~Bubeck \emph{et~al.}, ``The best of both worlds: stochastic and adversarial
  bandits,'' in \emph{Conference on Learning Theory}, 2012, pp. 1--23.

\bibitem{og}
R.~Kl{\'\i}ma \emph{et~al.}, ``Combining online learning and equilibrium
  computation in security games,'' in \emph{International Conference on
  Decision and Game Theory for Security}.\hskip 1em plus 0.5em minus
  0.4em\relax Springer, 2015, pp. 130--149.

\bibitem{psc}
B.~Reynders, W.~Meert†, and S.~Pollin, ``Power and spreading factor control
  in low power wide area networks,'' in \emph{IEEE ICC}, 2017.

\bibitem{opd}
A.~Azari, M.~Masoudi, and C.~Cavdar, ``{Optimized Resource Provisioning and
  Operation Control for Low-power Wide-area IoT Networks},'' \emph{arXiv
  preprint arXiv:1804.09464}, 2018.

\end{thebibliography}

\end{document}